\documentclass[noshowpacs,amsmath,twocolumn,
aps,prb]
{revtex4-1}

\usepackage{setspace}
\usepackage{amsmath}
\usepackage{graphicx}
\usepackage[nearskip,margin = 0pt]{subfig}
\usepackage{subfig}

\usepackage{verbatim}
\usepackage{amsfonts}
\usepackage{amssymb}
\usepackage{epstopdf} 
\usepackage{xcolor}
\usepackage{color}
\DeclareGraphicsExtensions{.pdf,.eps,.png,.jpg,.mps} 
\usepackage{ragged2e}
\usepackage{hyperref}
\hypersetup{
    colorlinks=true,
    linkcolor=blue,
    citecolor=blue,
    filecolor=blue, 
    urlcolor=blue,
}
\usepackage{mathrsfs}
\def\figurename{\bf Fig.}
\usepackage{threeparttable}
\usepackage{enumitem}

\begin{document}

\title{A microcomb-empowered Fourier domain mode-locked LiDAR}

\author{Zhaoyu Cai,$^{1}$ Zihao Wang,$^{1}$ Ziqi Wei,$^{1}$ Baoqi Shi,$^{2,3}$ Wei Sun,$^{2}$ Changxi Yang,$^{1}$ Junqiu Liu,$^{2,4}$ and Chengying Bao$^{1,*}$ \\
$^1$State Key Laboratory of Precision Measurement Technology and Instruments, Department of Precision Instruments, Tsinghua University, Beijing 100084, China.\\
$^2$International Quantum Academy, Shenzhen 518048, China.\\
$^3$Department of Optics and Optical Engineering, University of Science and Technology of China, Hefei, Anhui 230026, China.\\
$^4$Hefei National Laboratory, University of Science and Technology of China, Hefei 230088, China.\\
Corresponding author:  $^*$cbao@tsinghua.edu.cn 
}

\maketitle
\newcommand{\ts}{\textsuperscript}

\newcommand{\tsb}{\textsubscript}

%%%%%%%%%%%%%%%%%%%%%%%% Main Text %%%%%%%%%%%%%%%%%%%%%%%

{\bf Light detection and ranging (LiDAR) has emerged as an indispensable tool in autonomous technology. Among its various techniques, frequency modulated continuous wave (FMCW) LiDAR stands out due to its capability to operate with ultralow return power, immunity to unwanted light, and simultaneous acquisition of distance and velocity. However, achieving a rapid update rate with sub-micron precision remains a challenge for FMCW LiDARs. Here, we present such a LiDAR with a sub-10 nm precision and a 24.6 kHz update rate by combining a broadband Fourier domain mode-locked (FDML) laser with a silicon nitride soliton microcomb. An ultrahigh frequency chirp rate up to 320 PHz/s is linearized by a 50 GHz microcomb to reach this performance. Our theoretical analysis also contributes to resolving the challenge of FMCW velocity measurements with nonlinear frequency sweeps and enables us to realize velocity measurement with an uncertainty below 0.4 mm/s. Our work shows how nanophotonic microcombs can unlock the potential of ultrafast frequency sweeping lasers for applications including LiDAR, optical coherence tomography and sensing. %Besides LiDAR, our microcomb-based frequency calibration technique can also be useful to spectroscopy, optical coherent tomography, and to investigate lasing dynamics in ultrafast sweeping lasers.
}

Photonic integrated circuits (PICs) are playing an increasingly important role in LiDAR technology \cite{Rho_NN2021nanophotonics,Watts_Nature2013large,Watts_OL2017coherent,Wu_Nature2022large}. Besides beam steering, integrated microresonator frequency combs (microcombs) \cite{Kippenberg_NP2014,Kippenberg_Science2018Review,Bowers_NP2022integrated} can be invaluable light sources for LiDARs. Indeed, they have been used for dual-comb ranging \cite{Vahala_Science2018Range,Kippenberg_Science2018Range}, chaotic ranging \cite{Wang_NP2023breaking,Kippenberg_NP2023chaotic} and parallel frequency-modulated continuous wave (FMCW) ranging \cite{Kippenberg_Nature2020massively}. In these reports \cite{Vahala_Science2018Range,Kippenberg_Nature2020massively,Wang_NP2023breaking,Kippenberg_NP2023chaotic,Kippenberg_Science2018Range}, microcombs are usually sent to targets for ranging. Since the power of a microcomb is relatively low, high power amplification is frequently used \cite{Kippenberg_Nature2020massively,Wang_NP2023breaking,Kippenberg_NP2023chaotic,Kippenberg_Science2018Range}. In this work, we use an integrated silicon nitride (Si$_3$N$_4$) microcomb as a frequency ruler at the local site, as opposed to sending to targets, to calibrate a frequency sweep laser for FMCW ranging \cite{behroozpour2017lidar,lihachev_NC2022low,snigirev_Nature2023ultrafast,Kippenberg_Nature2020massively,rogers2021universal,qian_NC2022video,wang2024high,behroozpour2016electronic}. It not only avoids the power issue, but also leverages the large line spacing of microcombs to calibrate frequency sweeping with ultrahigh chirp rates for ranging. This is because the highest chirp rate that can be calibrated by a comb is limited to $f_{r}^2/4$ ($f_{r}$ is the comb repetition rate or line spacing) \cite{Newbury_STJQE2011characterizing}.

Calibrating frequency sweeping lasers by optical frequency combs has been widely used for ranging and spectroscopy  \cite{Kippenberg_NP2009frequency,Newbury_NP2010fast,Newbury_STJQE2011characterizing,Newbury_OL2013comb,VTC_OE2021frequency,Liu_PR2024frequency,Vahala_Science2019vernier,Zhang_OL2021nonlinear}.  Compared to interferometer or cavity based frequency calibration methods, combs provide higher accuracy and better immunity to environment fluctuations \cite{Newbury_NP2010fast,Newbury_OL2013comb}. Typically, femtosecond laser combs with repetition rates of hundreds of MHz were used for calibration, which limits the chirp rate to a few PHz/s \cite{Newbury_STJQE2011characterizing}. Soliton microcombs have been used for laser frequency calibration, but the measured chirp rate was below 12.5 THz/s \cite{Vahala_Science2019vernier,Zhang_OL2021nonlinear}. Recently, frequency sweep lasers with chirp rates above hundreds of PHz/s have been demonstrated using Fourier domain mode-locked (FDML) lasers\cite{Fujimoto_OE2006fourier,grill2022towards,Jalali_NP2020time} and micro-electro-mechanically tuned vertical-cavity surface emitting lasers (MEMS-VCSELs) \cite{Fujimoto_JLT2015wideband,chen2022high}. %The importance of these ultrahigh chirp rate  lasers can be understood by the example of FMCW LiDAR. 
These lasers are compelling because their ultrahigh chirp rates and broad bandwidths can boost the FMCW ranging speed, resolution and precision. However, these ultrahigh chirp rate lasers generally sweep in a nonlinear way. Their frequency calibration by microcombs can greatly enhance the functionality in LiDAR, optical coherence tomography (OCT) \cite{Fujimoto_OE2006fourier}, sensing \cite{jung2008characterization} and spectroscopy \cite{kranendonk2007high}. 

Here, a Si$_3$N$_4$ microcomb \cite{Liu_PR2023foundry} calibrates the frequency of a broadband FDML laser to reach a normalized ranging precision of 0.27 nm $\cdot \sqrt{\rm s}$, an order of magnitude higher than previously reported FMCW LiDARs \cite{Newbury_OL2013comb}. %and surpasses most dual-comb ranging results \cite{Newbury_NP2009rapid,Wu_Engineering2018}. 
The highest chirp rate calibrated is 320 PHz/s, which is two orders of magnitude higher than previous reports \cite{Newbury_STJQE2011characterizing,Newbury_OL2013comb} and highlights the unique advantage of large line spacing microcombs. Additionally, there have been assertions that nonlinear frequency sweeps can obscure velocity measurements, even after calibration \cite{Wu_OE2019laser}. We show that large chirp rates can mitigate the impact of frequency sweep nonlinearity, and our system achieves velocity measurements with an uncertainty below 0.4 mm/s. Our work also reveals the frequency sweeping dynamics of an FDML laser with an unprecedented bandwidth and resolution. 

%For higher chirp rates, frequency combs with a larger $f_r$ is needed. This makes PIC-based Kerr microcombs \cite{Kippenberg_Science2018Review,Bowers_NP2022integrated} a natural candidate for the task. Here, we use a silicon nitride (Si$_3$N$_4$) microcomb with $f_r$=50 GHz to calibrate an FDML laser with a chirp of 320 PHz/s for high precision FMCW ranging and velometry.

\begin{figure*}[t!]
\begin{centering}
\includegraphics[width=0.98\linewidth]{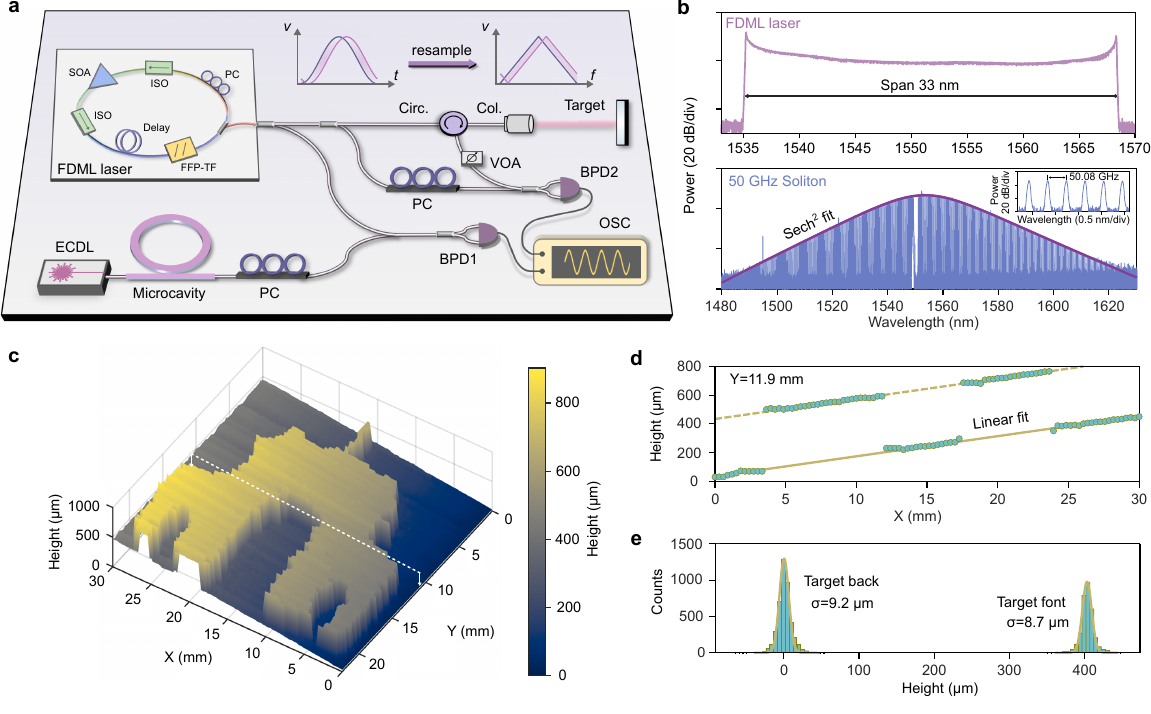}
\captionsetup{singlelinecheck=off, justification = RaggedRight}
\caption{\textbf{Architecture of FDML LiDAR and its application in 3D imaging}. \textbf{a,} Experimental setup of the FDML laser, which is calibrated by an integrated Si$_3$N$_4$ soliton microcomb for FMCW ranging.  
\textbf{b,} Optical spectra of the FDML laser and the 50 GHz soliton microcomb.
\textbf{c,} 3D imaging of a diffusive target representing the Tsinghua Gate, which was slightly tilted in the measurement.  
\textbf{d,} Measured height change of the target and its linear fits along the dashed line in panel \textbf{c}.  
\textbf{e,} Distribution of the height after subtracting the solid linear fit.
} 
\label{Fig1}
\end{centering}
\end{figure*}

\begin{figure*}[t!]
\begin{centering}
\includegraphics[width=0.98\linewidth]{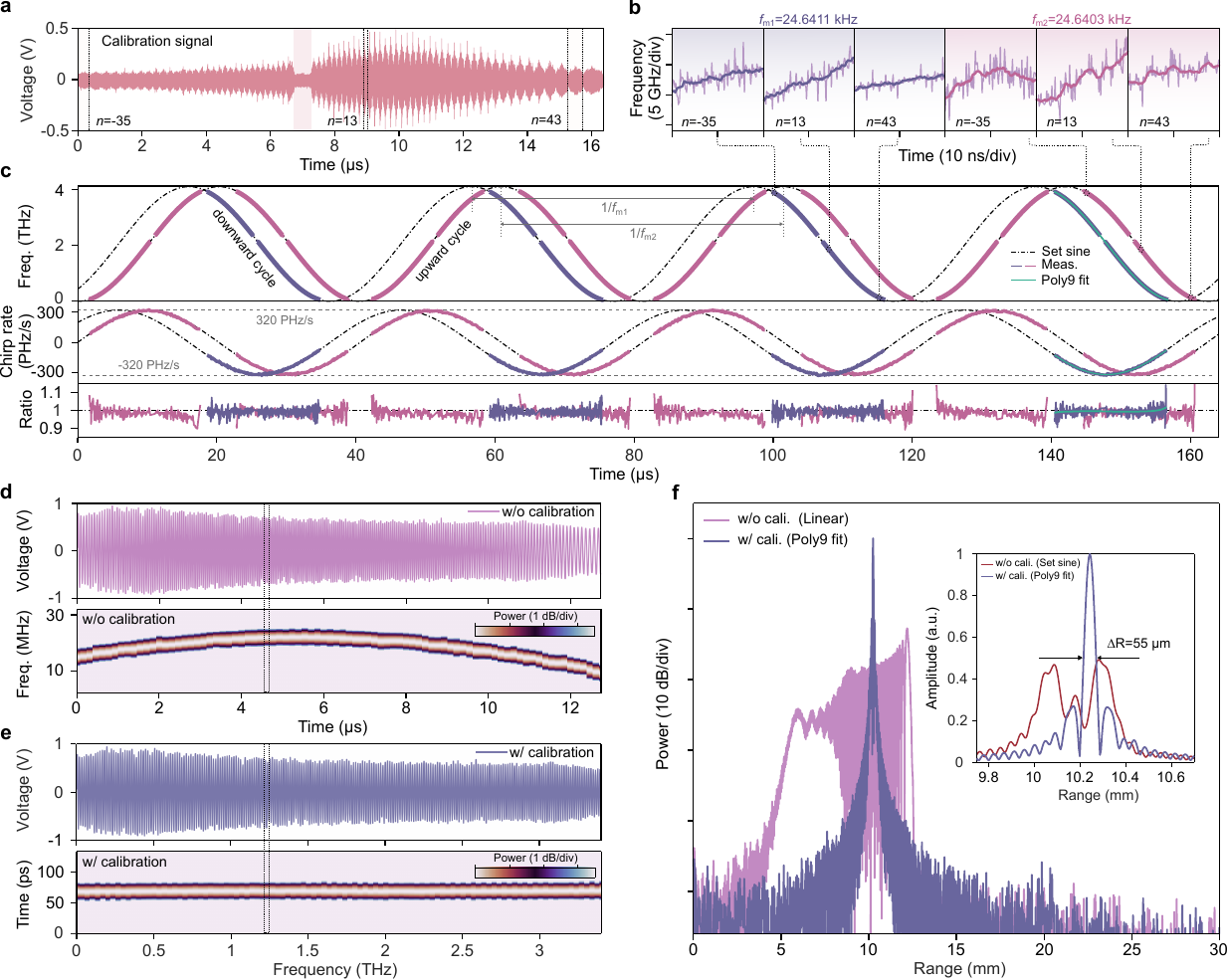}
\captionsetup{singlelinecheck=off, justification = RaggedRight}
\caption{\textbf{Microcomb-based frequency calibration for FMCW ranging}. \textbf{a,} Heterodyne beat signal between the microcomb and the FDML laser. The shaded region corresponds to the filtered pump and adjacent lines.  
\textbf{b,} Portion of the calibrated frequency sweep using three different lines under two drive frequencies $f_{m1}$ and $f_{m2}$. The light curves are calculated from the direct retrieved phase change, while the dark curves are calculated from phase change smoothed in a 2 ns window.
\textbf{c,} Top: calibrated frequency change after phase smoothing within the whole frequency sweep span. The green curve is a 9th-order polyfit of the frequency sweep signal. Middle: calibrated chirp rate of the FDML laser. Bottom: ratio between the calibrated chirp rates and chirp rates derived from an assumed sinusoidal frequency sweep.
\textbf{d,} Measured FMCW ranging signal in the time domain and its instantaneous spectra within a 0.15 $\mu$s window (dashed lines).  
\textbf{e,} Resampled FMCW signal and the instantaneous spectra.  
\textbf{f,} FMCW ranging output when using the direct measured signal and the resampled data. The inset shows the linear ranging outputs using the microcomb frequency calibration and an assumed sine-wave for resampling.} 
\label{Fig2}
\end{centering}
\end{figure*}

\begin{figure*}[t!]
\begin{centering}
\includegraphics[width=0.98\linewidth]{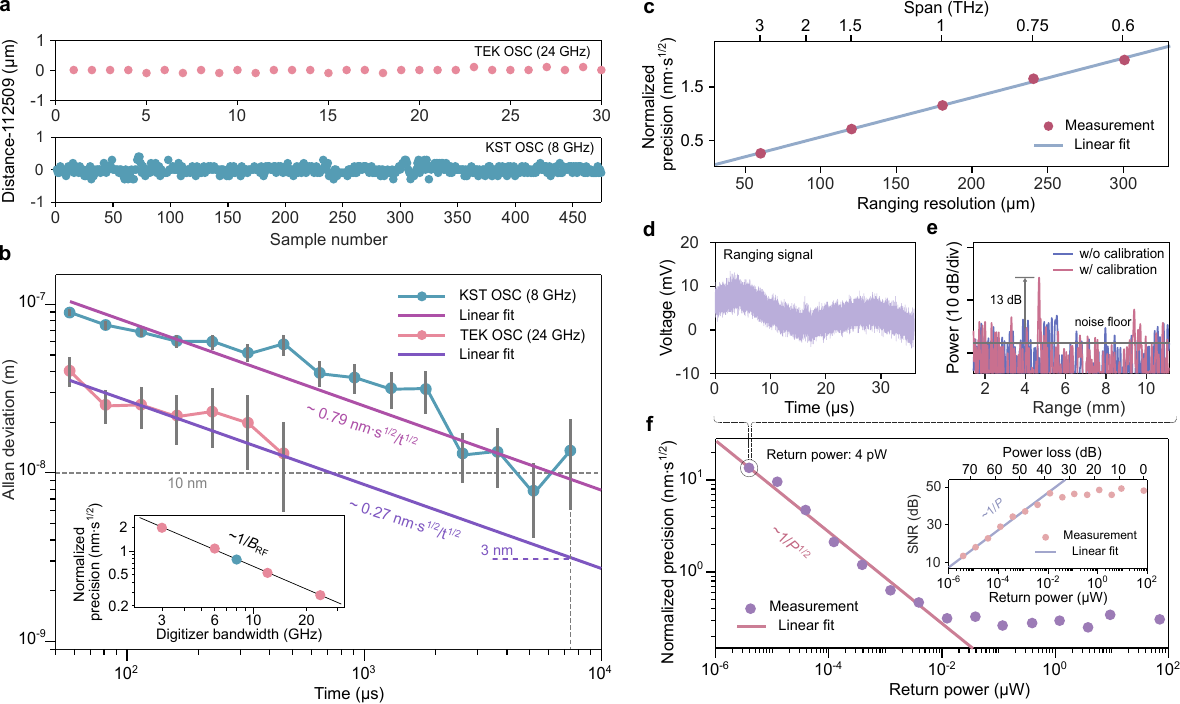}
\captionsetup{singlelinecheck=off, justification = RaggedRight}
\caption{\textbf{Precison of the FDML LiDAR.} \textbf{a,} Distance in multiple measurements when the calibration signal was recorded by a 33 GHz oscilloscope and a 8 GHz oscilloscope. The 8 GHz oscilloscope has a larger memory depth and recorded data for a longer time. 
\textbf{b,} Allan deviation of the measured distance, which scale as $t^{-1/2}$ with the averaging time $t$. Sub-10 nm precision is achieved in less than 10 ms. The inset shows normalized precision scales as $1/{B_{\rm RF}}$ with $B_{\rm RF}$ being the digitizer bandwidth.
\textbf{c,} Normalized precision scales linearly with the ranging resolution.  
\textbf{d,} Time domain FMCW ranging signal with a return power of 4 pW. 
\textbf{e,} FMCW ranging output using the direct time domain signal and the microcomb-calibrated signal with a 4 pW return power. 
\textbf{f,} Measured normalized ranging precision with different return powers (local oscillator power was 75 $\mu$W). The normalized precision scales as $P^{-1/2}$ for return powers $P<$10 nW. The inset shows SNR under different return powers.
} 
\label{Fig3}
\end{centering}
\end{figure*}

\begin{figure*}[t!]
\begin{centering}
\includegraphics[width=0.98\linewidth]{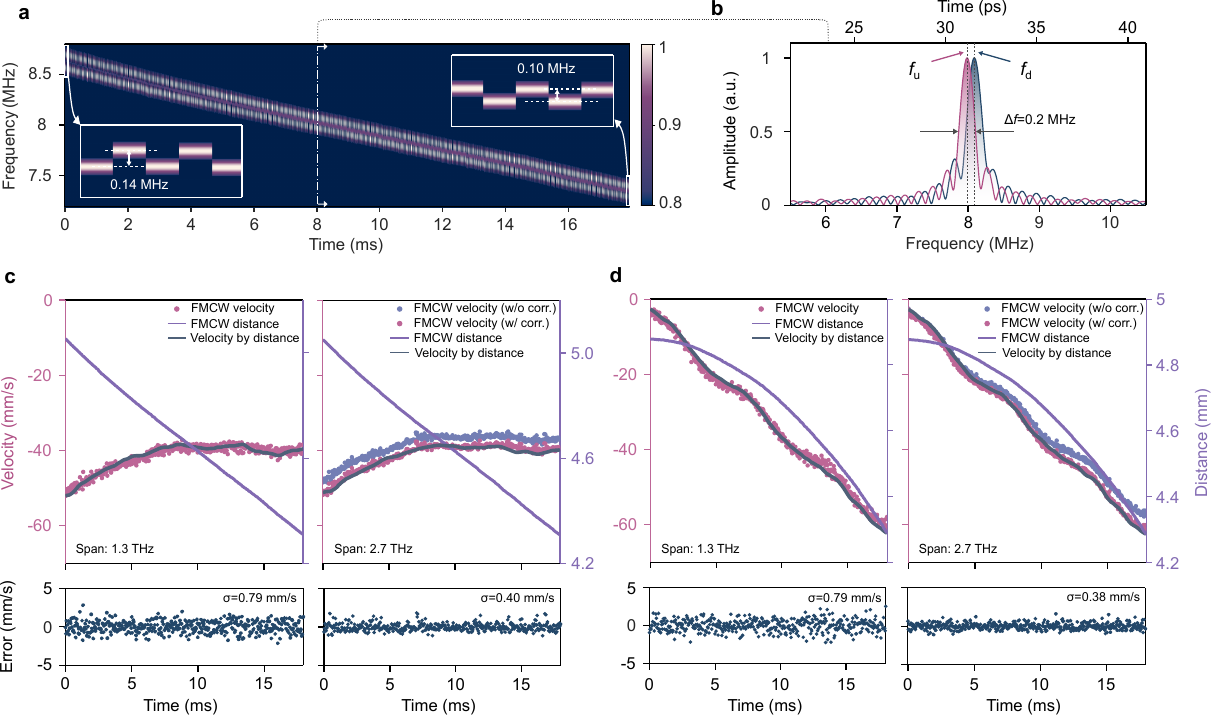}
\captionsetup{singlelinecheck=off, justification = RaggedRight}
\caption{\textbf{Velocity measurements using the FDML LiDAR.} 
\textbf{a,} Continuous ranging spectra featuring a decrease in mean frequency and frequency difference, indicating a deceleration motion of the target.
\textbf{b,} Ranging spectra at 8 ms of panel \textbf{a} in a linear scale (using a 1.3 THz lasing bandwidth). 
\textbf{c, d} FMCW measured velocity and velocity calculated from the measured distance. The bottom panels are residual error between the FMCW velocity and its polyfit. When using 2.7 THz lasing bandwidth for measurement, a correction factor should be included, due to the strong frequency sweep nonlinearity. Data were measured by the 8 GHz Keysight oscilloscope. 
%\textbf{d,} Velocity measurement results of an acceleration from rest performed by the 8 GHz oscilloscope. Velocities as low as near 0 mm/s can still be measured due to the high velocity measurement precision.
%\textbf{e,} Velocity measurement results performed by the 33 GHz oscilloscope, featuring a shorter record  time but a slightly higher precision. 
}
\label{Fig4}
\end{centering}
\end{figure*}

\noindent \textbf{LiDAR system and 3D imaging.} Our LiDAR system is illustrated in Fig. \ref{Fig1}a. The FDML laser works by matching the modulation frequency ($f_{m}$) of a fibre Fabry-Perot tunable filter (FFP-TF) with the free-spectral-range (FSR) of the cavity \cite{Fujimoto_OE2006fourier} (see Methods). In such a way, all the lasing frequencies can be stored in the cavity, which avoids the lasing buildup process and enables ultrahigh sweep rates. By carefully managing the dispersion, a lasing bandwidth of 33 nm at a rate of 24.6 kHz can be attained (Fig. \ref{Fig1}b). Since the tunable filter is driven by a sine-wave, the FDML frequency sweeps in a nonlinear way. To facilitate FMCW ranging, which works by homodyne beat between a frequency sweep laser and its delayed replica from the target (Fig. \ref{Fig1}a), we use a Si$_3$N$_4$ soliton microcomb with $f_{r}$=50.08 GHz, whose spectrum is shown in Fig. \ref{Fig1}b, to calibrate the FDML frequency in full-time (see Methods).

To showcase the capability of the microcomb-empowered FDML LiDAR, a 3D imaging of the Tsinghua Gate made of diffusive aluminum is presented in Fig. \ref{Fig1}c. We scanned the target mechanically for imaging and the received power loss from the target can be as high as 70 dB. The beam steering can be implemented by an optical phase array \cite{Watts_OL2017coherent} in the future. We deliberately tilted the target to examine the ability of our LiDAR to measure small height changes. Taking the dashed horizontal slice as an example, the measured height along the slice is shown in Fig. \ref{Fig1}d, showing a good linearity. By subtracting the solid linear fit in Fig. \ref{Fig1}d from the measured height, we have the height distribution shown in Fig. \ref{Fig1}e. The standard deviation ($\sigma$) is about 9 $\mu$m, which mainly results from the surface roughness of the target. An independent ranging linearity measurement shows a residual error with $\sigma$=0.2 $\mu$m (Extended Data Fig. 1).

\noindent \textbf{Frequency calibration in FMCW ranging.} Frequency calibration is essential in the above measurement. We heterodyne beat the FDML laser with the microcomb on a balanced photodetector (BPD) for this calibration. The output was measured by an oscilloscope with a bandwidth of 33 GHz (Tektronix MSO 73304DX). To avoid recording the beat signal with two adjacent lines simultaneously, we set the bandwidth of the oscilloscope to 24 GHz. The beat signal can be frequency-divided to relax the requirement on the digitizer bandwidth. An example of the heterodyne beat signal with over 70 microcomb lines is shown in Fig. \ref{Fig2}a, where the shaded region corresponds to the filtered pump and adjacent lines (Methods). Heterodyne beat with a CW laser has been used to study the phase stability of an FDML laser \cite{grill2022towards}. Our microcomb enables real-time phase measurements with a much larger bandwidth than the CW laser scheme. 

By retrieving the phase of the recorded signal via the Hilbert transform, the instantaneous frequency of the FDML laser can be measured (Methods and Extended Data Fig. 2). We show portions of the instantaneous frequency change measured by 3 different comb lines in Fig. \ref{Fig2}b ($n$ is the comb line number with respect to the pump). For two modulation frequencies $f_{m1}$=24.6411 kHz and $f_{m2}$=24.6403 kHz, the instantaneous frequencies change distinctively. For $f_{m1}$, there are abrupt frequency jumps at a frequency of about 0.8 GHz, while the frequency spurs occur almost irregularly for $f_{m2}$ (Extended Data Fig. 3). By smoothing the measured phase in a 2 ns time window, the spurs can be averaged and the smoothed frequency sweeps are plotted as the dark blue and pink curves in Fig. \ref{Fig2}b. The smoothed frequency sweeps over the full FDML lasing span are plotted in Fig. \ref{Fig2}c. Note that FDML lasing exists in both the upward and the downward sweep cycles for $f_{m2}$, while it only exists in the downward cycle for $f_{m1}$. Similar lasing asymmetry was reported in ref. \cite{jeon2008characterization}. The chirp rates were also calculated and plotted in Fig. \ref{Fig2}c. In general, both the laser frequency and the chirp rate track sine-waves set at $f_{m1,2}$. When normalizing the calibrated chirp rates by the assumed sine-waves, we can observe the chirp rate actually deviates from the set sweep, and the deviation is stronger at the edges of the frequency sweep span. Frequency sweep manner is also observed to be different in different modulation cycles. Microcombs can be used as a tool to better understand the ultrafast sweeping FDML or MEMS-VCSEL lasing dynamics in future work. %with unprecedented bandwidth and accuracy.

Since the instantaneous frequency change is steadier for $f_{m1}$, we used this frequency for the ranging experiments. The temporal ranging signal recorded by BPD2 (see Fig. \ref{Fig1}a) is shown in Fig. \ref{Fig2}d. As the output of BPD2 was low-pass filtered by a 100 MHz filter, the frequency spurs in Fig. \ref{Fig2}b do not impact this signal. When gating the signal by a 0.15 $\mu$s window (dashed lines in Fig. \ref{Fig2}d) to look into the instantaneous spectra, time-varying frequency is observed. Since the smoothed frequency sweep in Fig. \ref{Fig2}b is still influenced by the frequency spurs, we further used the 9th-order polyfit to fit the frequency sweep (green curves in Fig. \ref{Fig2}c). Then, the fitted frequencies were used to resample the FMCW ranging signal in the frequency domain (Methods) \cite{Newbury_OL2013comb}. The resampled instantaneous oscillation `frequency' becomes uniform (Fig. \ref{Fig2}e). By Fourier transforming the resampled signal, we obtained the ranging signal in Fig. \ref{Fig2}f. A sharp peak centered at 10.2 mm with a resolution of 55 $\mu$m can be observed. When zooming in the peak in a linear scale, a transform-limited signal can be observed. %, whose resolution is \textcolor{magenta}{50 $\mu$m (60 $\mu$m/1.2)} and is limited by the FDML laser frequency sweep bandwidth (3 THz). 
In contrast, if we Fourier transform the time-domain signal directly, the ranging spectrum is broad and meaningless. As the FDML laser was driven by a sine-wave, we also assumed the FDML lasing frequency changes in a sinusoidal way with the frequency set at $f_{m1}$ and the starting/ending frequency read from the optical spectrum. When resampling the time-domain signal by this assumed sine-wave, the Fourier transformed signal is still broad (inset of Fig. \ref{Fig2}f). Therefore, it is critical to have the FDML laser precisely calibrated for ranging, and we have an ultrahigh chirp rate of 320 PHz/s calibrated for FMCW ranging. % and our microcomb enables frequency calibration for chirp rates up to 320 PHz/s (Fig. \ref{Fig2}b). 

\noindent \textbf{High precision FMCW ranging.} Precise frequency calibration and broad bandwidth of the FDML laser enable sub-10 nm precision. We show the continuously measured distance in Fig. \ref{Fig3}a. Since the 33 GHz oscilloscope has a limited memory depth, we also used a 8 GHz oscilloscope (Keysight DSOS804A) to record the FMCW ranging signal in a longer time. The corresponding Allan deviation is plotted in Fig. \ref{Fig3}b. The precision scales as 0.27 nm$\cdot \sqrt{\rm s}/t^{1/2}$ and 0.79 nm$\cdot \sqrt{\rm s}/t^{1/2}$ ($t$ is the measurement time), respectively. Since the 8 GHz oscilloscope only calibrate the FDML laser in about a third of the frequency span, the precision is worse than the 33 GHz oscilloscope data. Despite the limited bandwidth, the results obtained by the 8 GHz oscilloscope still reaches sub-10 nm precision in 5 ms. Furthermore, it shows the Allan deviation keeps converging in at least 8 ms. Hence, we believe that our FMCW LiDAR can reach a precision of 3.0 nm at 8 ms, once sufficiently long data are recorded. 

We further low-pass filtered the 33 GHz oscilloscope recorded data by different bandwidth $B_{\rm RF}$ to analyze the ranging precision. The normalized precision scales as $1/{B_{\rm RF}}$ and the 8 GHz oscilloscope data also aligns with the scaling (inset of Fig. \ref{Fig3}b). Therefore, only the calibrated frequencies contribute to the improvement of the ranging precision. This means full-time frequency calibration is more advantageous than only using combs for tick-like calibration \cite{Kippenberg_NP2009frequency,Zhang_OL2021nonlinear}.

As noted in ref. \cite{Newbury_Nature2022time}, ranging precision is proportional to $\Delta R/\sqrt{\rm SNR}$, where $\Delta R$ is the resolution and SNR is the signal-to-noise ratio in the power spectrum of the ranging signal. By selecting different frequency spans of the calibrated signal in Fig. \ref{Fig2}e, we derive the normalized precision under different resolutions (Fig. \ref{Fig3}c). As expected, a linear relationship with ranging resolution $\Delta R$ is observed. 
To further examine the dependence on SNR, we attenuated the received power to measure the ranging precision (see Fig. \ref{Fig1}a). When reducing the power to 4 pW (that is about 600 photons in the used downward sweep cycle and 72 dB loss in the signal arm), the recorded signal in the time domain almost shows no interferometric features. In the `ranging' domain, a peak with SNR of 13 dB can still be observed. No such peak can be observed when the frequency sweep is not calibrated. The normalized precision in this case is 13 nm$\cdot\sqrt{\rm s}$. When increasing the received power, the normalized Allan deviation decreases in an inverse square-root way, as the SNR is determined by the product of the local and received powers (Fig. \ref{Fig3}f and its inset). Ranging precision and SNR saturate for received power above 10 nW, which may be attributed to the spurious reflections
in the fibres  \cite{Newbury_OL2013comb}. Once the parasitic reflections dominate the noise floor, increasing received power no longer improves the SNR and the precision is clamped.    

%The system can still measure the range when the power reflected from the target is ultralow. When the return power is as low as 3.9 pW, the ranging signal recorded by BPD2 in Fig. \ref{Fig1}a is shown in Fig. \ref{Fig3}c. It is difficult to distinguish the sinusoidal signal from this time domain signal due to the ultralow signal-noise-ratio (SNR). As shown in Fig. \ref{Fig3}d, no obvious peaks are seen after Fourier transformation without calibration. After calibration, a peak with a SNR of 13 dB can be seen. The peak center of 4.7 mm corresponds to the measurement distance. Fig. \ref{Fig3}e shows the precision in the 40 $\mu$s update interval as a function of return power when the reference power is 75 $\mu$W. The precision does not improve beyond 10 nW return power because the SNR clamps at 50 dB (Fig. \ref{Fig3}e inset) due to the calibration precision. Apart form the SNR, the ranging resolution also contributes to the precision. The precision with 40 $\mu$s measurement time as a function of frequency sweep bandwidth is plotted in Fig.\ref{Fig3}f, showing a good linearity as ref.\cite{Naim_AO2011high} and ref.\cite{Newbury_OL2014Speckle} estimated.

\begin{figure}[t]
%\begin{centering}
\includegraphics[width=0.95\linewidth]{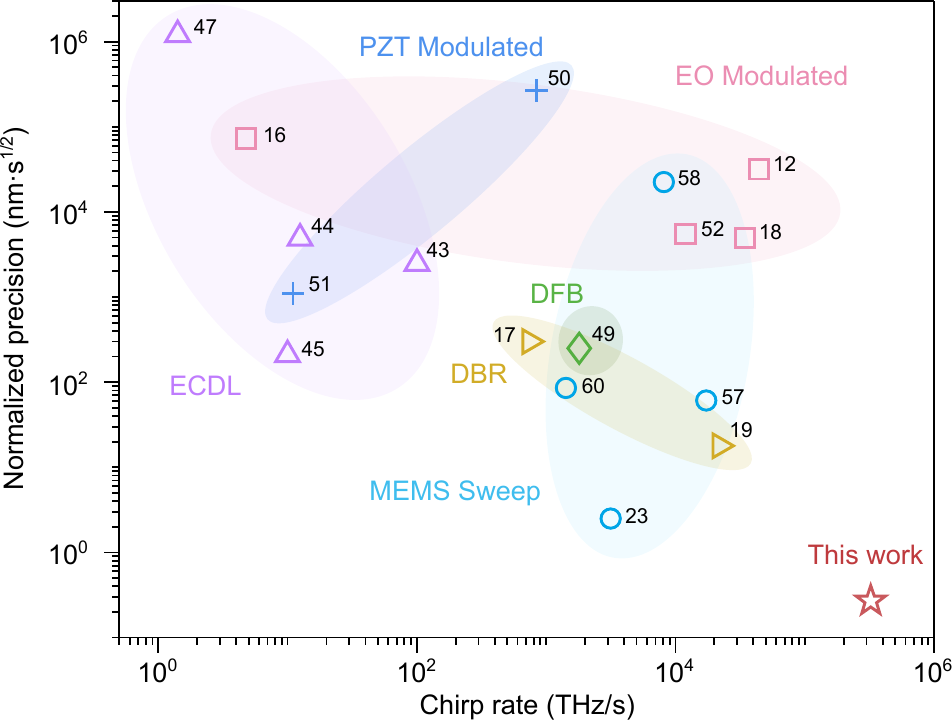}
\captionsetup{singlelinecheck=off, justification = RaggedRight}
\caption{{\bf FDML LiDAR performance in comparison with other reports.} Our system has an excellent combination of normalized precision and chirp rate. 
%The prefix S represents references in the Supplementary Information.
}
\label{Fig5}
%\end{centering}
\end{figure}

\noindent \textbf{Velocity measurements.}
%The ability to measure velocity and distance simultaneously is a unique advantage for FMCW LiDARs. 
Doppler frequency shift from a moving target causes frequency split in the FMCW ranging spectra for the upward and downward sweep cycles. %The splitted frequency peaks can be denoted as $f_{\rm u}$ and $f_{\rm d}$, respectively. Then, velocity along the light propagation direction is derived as $\lambda_{\rm c}(f_{\rm u}-f_{\rm d})/4$, where $\lambda_{\rm c}$ is the carrier wavelength \cite{Kippenberg_Nature2020massively}. 
With frequency calibration and data resampling, the ranging spectrum has a time-axis (Figs. \ref{Fig2}e, f) rather than a frequency-axis. The time-axis should be converted back to a frequency-axis to locate the two split frequency peaks $f_{u}$ and $f_{d}$ for velocity measurements. Artifacts can arise in the conversion with nonlinear frequency sweeps. Large chirp rates help mitigate the artifacts (Methods).

In this experiment, we set the drive frequency at $f_{m2}$ to have FDML lasing in both sweep cycles. The continuous ranging spectra measured from a moving target is shown in Fig. \ref{Fig4}a (recorded by the 8 GHz oscilloscope). The inset clearly shows the frequency split and the deceleration of the target.  The ranging spectra at 8 ms are plotted in Fig. \ref{Fig4}b, indicating that both the upward and the downward cycles are calibrated to have transform-limited ranging signals. We use a relatively sweep bandwidth of 1.3 THz for this measurement so as to lower the frequency sweep nonlinearity, and the time-axis is transferred to the frequency-axis by simply multiplying the overall chirp rate (Methods). 

We show the distance and velocity measured in real-time in Figs. \ref{Fig4}c, d, corresponding to a deceleration and an acceleration scenario, respectively. The FMCW measured velocity agrees well with the velocity calculated from the measured distance. The residual error between the FMCW measured velocities and their polyfit has a standard deviation as small as 0.79 mm/s. This velocity uncertainty can be further reduced to 0.40 mm/s by using a 2.7 THz bandwidth for measurement. With this broader bandwidth, the frequency sweep nonlinearity becomes larger and a correction factor should be included (Figs. \ref{Fig4}c, d), see Methods and Extended Data Fig. 4 for details. The velocity uncertainty is fairly low considering the ultrahigh chirp rate, which extends the range of measurable velocity but adds difficulty in measuring velocity precisely (Methods and Extended Data Table 1).

\noindent \textbf{Discussions.} Our work establishes a technique to tame ultrafast frequency sweep lasers (theoretically up to chirp rates of hundreds of EHz/s) for FMCW LiDAR with low return power. A PIC-based microcomb calibrates a broadband FDML laser with ultrahigh chirp rates to reach the best normalized precision for FMCW LiDARs at an update rate of 24.6 kHz (Fig. \ref{Fig5} and Extended Data Table 1). %To our knowledge, our system represents the highest chirp rate and the best normalized precision for FMCW LiDARs (see Fig. \ref{Fig5} and Extended Data Table 1). 
FDML lasers with lasing bandwidth exceeding 20 THz and MHz update rates have been demonstrated \cite{kolb2018high}. Together with a dispersion-engineered, broadband microcomb, it may further improve the ranging precision and measurement speed. %Ambiguity range is another important parameter for LiDARs.
Although the absolute distance in FMCW LiDARs can be adjusted by tuning the length of the reference arm, the measurement range is limited by the linewidth of the frequency sweeping laser. Our FFP-TF has a relatively broad bandwidth, which limits this range to about ten centimetre. Hence, the current system may be more suitable for fast 4D imaging (including velocity) applications. By optimizing the dispersion management and controlling $f_{m}$ precisely, the coherence length of an FDML laser can reach metre-scale \cite{pfeiffer2017analysis}. This requires meticulous control of the laser configuration. Frequency sweep dynamics revealed by the microcomb can be used in turn to optimize the laser condition. MEMS-VCSEL lasers with ultrahigh chirp rates and narrow linewidths can pair with microcombs similarly to extend the measurement range to hundreds of metres \cite{Fujimoto_JLT2015wideband}. Finally, our work can guide precise FMCW velocity measurement with frequency sweep nonlinearity. 

\vspace{3 mm}
\noindent{\bf Methods}

{\small
\noindent \textbf{Fourier domain mode-locked laser.} The FDML laser consists of a FFP-TF (MOI-FFP-TF2-1520-1570-7.5G2000) with a finesse of 2000 and an FSR of 135 nm, a semiconductor optical amplifier (Thorlabs SOA117S), a fibre delay of 8.1 km consisting of a 0.9 km dispersion compensation fibre (DCF) and a 7.2 km single mode fibre (SMF). Two optical isolators were used to ensure unidirectional operation and a polarization controller (PC) was used to optimize the lasing polarization. A pulse shaper (Finisar WS 1000B) was inserted for finer dispersion compensation. The filter was driven by a sinusoidal signal around 24.6 kHz, and the output was derived from the 30\%-port of a 70/30 coupler. The output power of the laser was about 1.3 mW, which is mainly limited by the damage power of the filter.

\vspace{1 mm}

\noindent \textbf{Silicon nitride soliton microcomb.} The foundry manufactured Si$_3$N$_4$ microresonator has a dimension of 0.81$\times$2.2 $\mu$m \cite{Liu_PR2023foundry}. The intrinsic and loaded Q-factors of the sample are 9.8 million and 6.4 million, respectively. An external cavity diode laser (ECDL, Toptica CTL1550) was amplified to pump the microresonator with an on-chip power of 200 mW. To mitigate the thermal instability, we used a single-sideband modulator (SSBM) based fast laser sweep technique to initiate the soliton generation \cite{Papp_PRL2018,liu2023mitigating}. To avoid the influence of the residual sidebands from the SSBM, we suppressed the frequencies near the pump by a notch filter with a bandwidth of 200 GHz (4 lines). The heterodyne beat between the soliton microcomb and the FDML laser was registered by a BPD (Finisar BPDV2150R) with a bandwidth of 43 GHz, whose output was amplified by a low noise amplifier.  

\vspace{1 mm}

\noindent \textbf{Phase retrieval and frequency calibration.} %A perfectly stable comb outputs a series of equally spaced pulses in the time domain, $t$, $E_{\rm c}(t)= \sum_{m} E_{\rm c0}\left(t-mT_{\rm r}\right){\rm exp}\left(im\theta_{\rm ceo}\right)$, where $m$ is the pulse number, $T_{\rm r}=1/f_{r}$ is the period and $\theta_{\rm ceo}=2\pi f_{\rm ceo}/f_{r}$ is the carrier envelope offset phase shift. For FDML laser field $E_{\rm L}(t)=E_{\rm L0}{\rm exp}\left[i\varphi_{\rm L}(t)\right]$, the calibration signal voltage $V(t_{\rm s})\propto E_{\rm L0}{\rm exp}\left[i\varphi_{\rm L}(t_{\rm s})-im\theta_{\rm ceo}\right]$ is sampled at discrete times $t_{\rm s}=mT_r$ when approximating the short pulses as delta functions. 
%The instantaneous frequency of $V(t_{\rm s})$, that is heterodyne frequency, is $f_{\rm h}\left(t_{\rm s}\right)=f_{\rm L}\left(t_{\rm s}\right)-f_{\rm ceo}-nf_{r}$, where $f_{\rm L}\left(t_{\rm s}\right)=\dot{\varphi}_{\rm L}(t_{\rm s})/2\pi$ is the instantaneous frequency of FDML laser and $n$ is the label of the nearest comb line. When filtering the calibration signal $V(t_{\rm s})$ with a low pass filter with a bandwidth of $f_{r}/2$, the heterodyne frequency is 'folded' by $f_{r}/2$. We apply the Hilbert transform to the calibration signal to retrieve the instantaneous phase and heterodyne frequency $f_{\rm h}$ (more detailed signal processing is shown in Supplementary Information).  
The phase retrieval process is also described in Extended Data Fig. 2. We first separated the measured heterodyne beat signal into different segments (each lasting 4.6 ns). Then, we Fourier transformed the segmented data and filtered out the strong peaks by a bandwidth of 3.5 GHz (Extended Data Fig. 2b). The filtered spectrum was inverse Fourier transform back to the time domain. Phase of the filtered signal was retrieved by Hilbert transform (Extended Data Fig. 2c). The phase was further smoothed in a 2 ns window to yield the instantaneous frequency (Extended Data Figs. 2d, e). Then the heterodyne frequency results is unwrapped by the comb spacing $f_{r}$=50.08 GHz to yield FDML laser instantaneous frequency in the full sweep span (Extended Data Fig. 2f).

\vspace{1 mm}

\noindent \textbf{Resampling of the ranging signal.}  For a distance $R$, the round trip time delay is $\Delta t=2R/c$, where $c$ is the light velocity in the air. We can write the local and signal frequency sweep field as
$E_{\rm LO}(t)=E_1(t){\rm exp}\left[i\varphi(t)\right]$ and $E_{\rm sig}=E_2(t+\Delta t){\rm exp}\left[i\varphi(t+\Delta t)\right]$, where $E_{1(2)}$ is the laser amplitude, and $\varphi(t)$ include the phase change due to the sweeping frequency. Since resampling mainly deals with the phase of the measured time domain signal, we first omitted the possible amplitude variation in the laser field. Then, the normalized output of BPD2 in Fig. \ref{Fig1}a can be written as,

\begin{equation}
V_{r}(t)={\rm cos} \left[\varphi(t+\Delta t)-\varphi(t) \right].
\label{eq1}
\end{equation}
Since the FDML laser sweeps in a nonlinear way, the instantaneous frequency of $V_{r}(t)$ (from a static target) is time varying and can be written as,
\begin{equation}
\begin{aligned}
f_{b}(t)= & \frac{1}{2\pi} \frac{d\left[\varphi(t+\Delta t)-\varphi(t) \right]}{dt}=\nu(t+\Delta t)-\nu(t) \\
 & \approx \Delta t \cdot {\rm d}\nu(t)/{\rm d}t,
\end{aligned}
\label{eq2}
\end{equation}
where $\nu(t)$ is the instantaneous frequency of the FDML laser. Eq. \ref{eq2} holds when $\Delta t$ is relatively small. 
%where the approximation error can be neglected because $\Delta t \ll 1/f_m  (R\ll 6 \rm km) $. 

Thus, $\varphi(t+\Delta t)-\varphi(t)$ can be calculated as,
\begin{equation}
\begin{aligned}
\varphi(t+\Delta t)-\varphi(t)=2\pi \int_{0}^{t} f_{b}(\tau){\rm d}\tau   \approx 2\pi \Delta t\left[\nu(t)-\nu(0)\right].
\end{aligned}
\label{eq3}
\end{equation}
The ranging signal can be written as, 
\begin{equation}
V_{r}(t)={\rm cos}\left(2\pi \Delta t \left[\nu(t)-\nu(0) \right]\right)=\cos{\left(2\pi \Delta t \nu_{s}(t) \right)},
\label{eq4}
\end{equation}
where we define $\nu_{s}(t)\equiv\nu(t)-\nu(0)$. By replacing the time-axis with the frequency-axis, the signal $V_{r}(t)$ is resampled as $V_{r}(\nu_{s})$. Then, $\Delta t$ can be derived via Fourier transform versus $\nu_{s}$. In the Fourier transform, we added zeros, whose length is nine times of the recorded data length, to the resampled data so as to locate the peak and $\Delta t$ more accurately. 

\vspace{1 mm}

\noindent \textbf{Velocity measurement.} When the target is moving, the beating frequency $f_{b}(t)$ becomes, 

\begin{equation}
\begin{aligned}
f_{b}(t) \approx \Delta t \cdot {\rm d}\nu(t)/{\rm d}t + f_{\rm D},
\end{aligned}
\label{eqnDoppler}
\end{equation}
where $f_{\rm D}=v/\lambda_c$ and $v$ is the velocity being measured. Thus, the phase difference  $\varphi(t+\Delta t)-\varphi(t)$ becomes, 

\begin{equation}
\begin{aligned}
\varphi(t+\Delta t)-\varphi(t) \approx 2\pi \Delta t \nu_{s}(t) + 2\pi f_{\rm D}t.
\end{aligned}
\label{eqPhiwV}
\end{equation}
$\nu_{s}(t)$ is positive (negative) in the upward (downward) cycle. We focus on the positive frequencies in the following analysis and write the complex form of $V_{r}(t)$ as, 
\begin{equation}
V_{r}(t)=\underbrace{\exp{(i2\pi \Delta t|\nu_{s}(t)|)}}_{V_{ r1}}\underbrace{\exp{(\pm i 2\pi f_{\rm D}t)}}_{V_{r2}}.
\label{eqnComplex}
\end{equation}
The ranging spectrum is $\widetilde{V}_{r1} \otimes \widetilde{V}_{r2}$, where $\widetilde{V}_{r1 (r2)}$ is the spectrum for $V_{r1(r2)}$ and $\otimes$ is the symbol for convolution. In a linear sweep case, $\nu(t)=\nu(0)+\nu_1 t$, where $\nu_1$ is the chirp rate. Resampling is not needed, and the ranging spectrum is, 

\begin{equation}
\widetilde{V}_{r}(f)=\delta(|\nu_1|\Delta t) \otimes \delta(\pm f_{\rm D})=\delta(|\nu_1|\Delta t\pm f_{\rm D}).
\label{eqnLSpectrum}
\end{equation}
Therefore, the ranging peaks are shifted by $\pm f_{\rm D}$ (i.e., spaced by 2$f_{\rm D}$). 

In a nonlinear sweep case, %Fourier transform of the resampled signal $V_{r1}(\nu)$ yields $\widetilde{V}_{r1}(t)=\delta(\Delta t)$. However, the resampling causes $V_{r2}(\nu)$ to vary nonlinearly with $\nu$. 
we define the relationship that $\nu_{s}(t)\equiv{\nu_1t+\nu_{e}(t)}$, where $\nu_1=(\nu(T)-\nu(0))/T$ ($T$ is the ending time of frequency sweep) is the overall chirp rate and $\nu_{e}$ is an error frequency. Then, $V_{r}(t)$ is resampled as,

\begin{equation}
V_{r}(\nu_{s})=\underbrace{\exp{(i2\pi \Delta t|\nu_{s}|})}_{V_{r1}}\underbrace{\exp{(\pm i 2\pi f_{\rm D}\nu_{s}/\nu_1)}}_{V_{r2}}\underbrace{\exp{(\pm i 2\pi f_{\rm D}\nu_{e}/\nu_1)}}_{V_{r3}}.
\label{eqnVresample}
\end{equation}
Fourier transform versus $\nu_{s}$ yields the ranging spectrum as,

\begin{equation}
\widetilde{V}_{r}(t)=\delta(\Delta t) \otimes \delta(\pm f_{\rm D}/\nu_1))\otimes\widetilde{V}_{r3}(t)=\delta(\Delta t\pm f_{\rm D}/\nu_1)\otimes\widetilde{V}_{r3}(t).
\label{eqnVTime}
\end{equation}
When $f_{\rm D}\nu_{e}/\nu_1$ is small, $V_{r3}(\nu_{s})$ can be approximated as a constant ($\widetilde{V}_{r3}(t)=\delta(0)$). Then, $\widetilde{V}_{r}(t)$ still recovers $f_{\rm D}$ and velocity (1.3 THz case in Fig. \ref{Fig4}). Therefore, velocity measurement works better, when the velocity is low (small $f_{\rm D}$) and selecting a relatively small optical bandwidth (small $\nu_{e}$). Moreover, the ultrahigh chirp rate of FDML lasers (large $\nu_1$) is beneficial for velocity measurement with frequency sweep nonlinearity and data resampling. 

When $f_{\rm D}\nu_{e}/\nu_1$ is large, $\widetilde{V}_{r3}(t)$ may distort the velocity measurement. Since $\nu_{s}(t)$ has been measured accurately by the microcomb, $\widetilde{V}_{r3}(t)$ can be calculated numerically with some prior knowledge of $f_{\rm D}$, which can be attained by using a small frequency sweep bandwidth first. Then its influence can be corrected (Extended Data Fig. 4).

Eq. \ref{eqnVTime} shows uncertainty in $\Delta t$ is multiplied by $\nu_1$ when determining $f_{\rm D}$, which means measuring velocity precisely is challenging for large chirp rates. The largest $f_{\rm D}$ that can be measured is $\nu_1 \Delta t$; and the measurable velocity is limited to $v_{\rm max}=\lambda_c \nu_1 \Delta t$.

%\underbrace{e^{i(\omega(t)-\omega(0)) \Delta t}}_{\mathrm{V}_{r 1}} \underbrace{e^{i \omega_D t}}_{V_{r 2}}

%In this case, $(f_{\rm d}+f_{\rm u})/2$ bears the $\Delta t$ and distance results, and $(f_{\rm u}-f_{\rm d})/2$ bears the $f_{\rm D}$ and velocity results ($f_{\rm u}$ and $f_{\rm d}$ are denoted in Fig. \ref{Fig4}b). 

\vspace{3 mm}
\noindent \textbf{Data Availability}
The data that support the plots within this paper and other findings are available.

\vspace{1 mm}
\noindent \textbf{Code Availability}
The code that supports findings of this study are available from the corresponding author upon request.

%\vspace{1 mm}
\noindent \textbf{Acknowledgements}
We thank Prof. Qiang Liu and Prof. Yidong Tan at Tsinghua University for discussions and equipment loan. The silicon nitride chip used in this work was fabricated by Qaleido Photonics. This work is supported by the National Key R\&D Program of China (2021YFB2801200), by the National Natural Science Foundation of China (62250071, 62175127), by the Tsinghua-Toyota Joint Research Fund, and by the Tsinghua University Initiative Scientific Research Program (20221080069). J.L. acknowledges support from the National Natural Science Foundation of China (12261131503), Innovation Program for Quantum Science and Technology (2023ZD0301500), Shenzhen-Hong Kong Cooperation Zone for Technology and Innovation (HZQB-KCZYB2020050), and Shenzhen Science and Technology Program (RCJC20231211090042078).

\vspace{1 mm}

\noindent\textbf{Author Contributions} Z.C. led the experiments with assistance from Z. Wang., Z. Wei, and C.Y.; B.S., W.S. and J.L. prepared and characterized the Si$_3$N$_4$ chip. The project was supervised by C.B.

\vspace{1 mm}
\noindent \textbf{Competing Interests} The authors declare no competing interests. }

\bibliography{main}

\clearpage
\setcounter{figure}{0}
\renewcommand{\figurename}{\bf Extended Data Fig.}
\begin{figure*}[t]
\begin{centering}
\includegraphics[width=0.95\linewidth]{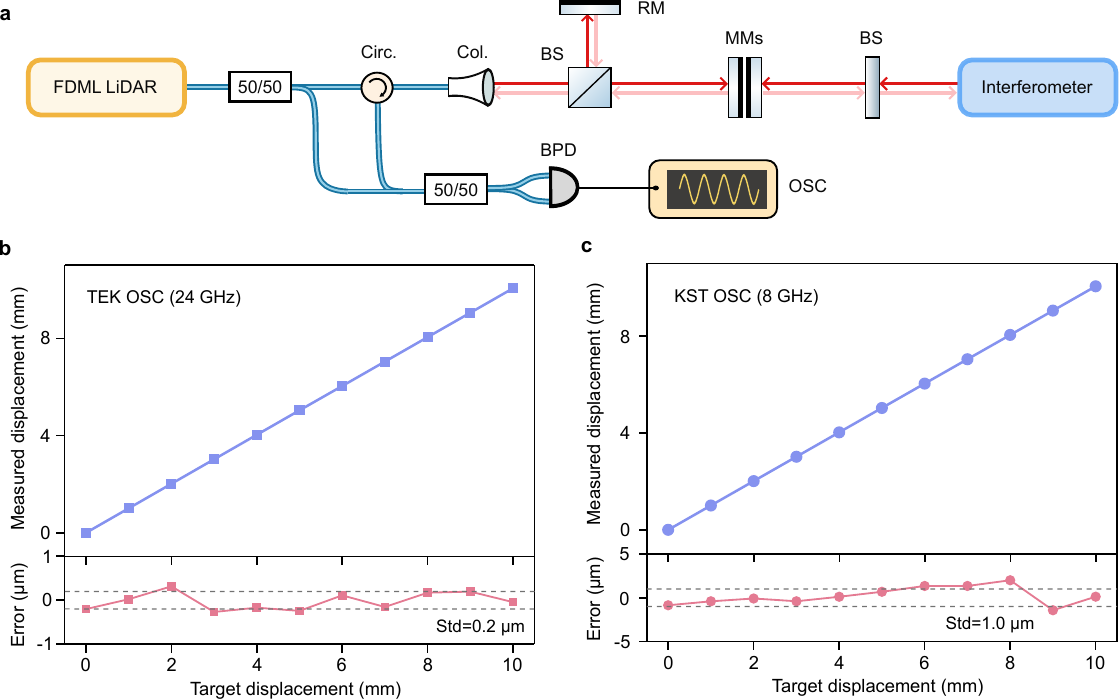}
\captionsetup{singlelinecheck=off, justification = RaggedRight}
\caption{{\bf Linearity measurement for the FDML LiDAR.} \textbf{a,} Experimental setup for the ranging linearity measurement. By mounting the mirrors on a linear translation stage, we characterized the linearity measurement accuracy of the FDML LiDAR. The movement of the target was calibrated by a laser interferometer. To cancel the fibre length fluctuations, we added a reference mirror in this setup too.
Circ.: circulator, Col.: collimator, BS: beam filter, RM: reference mirror, MMs: measurement mirrors, BPD: balanced photodetector. 
\textbf{b,} Top: FDML LiDAR measured distance change versus the laser interferometer calibrated distance change. The measurement data were recorded by the 33 GHz Tektronix oscilloscope (bandwidth set at 24 GHz). Bottom: The residual error between the FDML LiDAR and the interferometer measured results. The standard deviation is 0.2 $\mu$m.
\textbf{c,} The ranging linearity measurements recorded by the 8 GHz Keysight oscilloscope. The standard deviation of the residual error is 1.0 $\mu$m.}
\label{Extended_Fig1}
\end{centering}
\end{figure*}

\begin{figure*}[t]
\begin{centering}
\includegraphics[width=\linewidth]{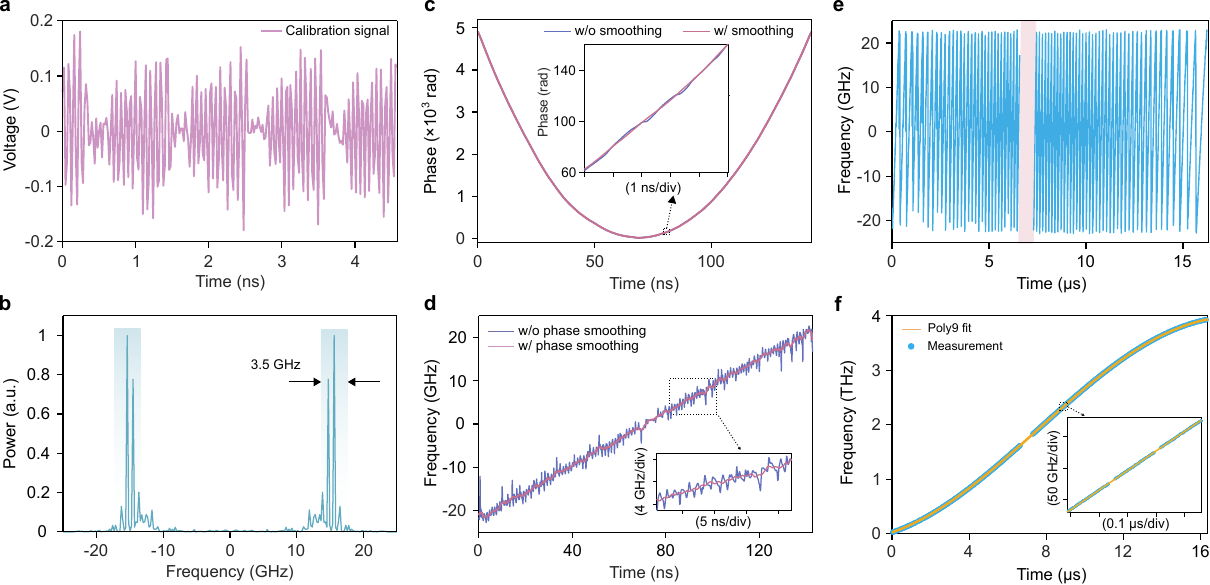}
\captionsetup{singlelinecheck=off, justification = RaggedRight}
\caption{{\bf Frequency calibration process.} \textbf{a,} We first selected a segment of the recorded calibration signal by a 4.6 $\mu$s window.
\textbf{b,} The segmented data was Fourier transformed and bandpass filtered (shaded boxes represent the passband). 
\textbf{c,} The bandpass filtered spectrum was inverse Fourier transformed back to the domain. The phase of the signal was retrieved by the Hilbert transform (blue curve). The retrieved phase was further smoothed in a 2 ns window (red curve). %The instantaneous phase obtained by Hilbert transforming the filtered calibration signal still shows small fluctuations and gets smoother with average filtering.
\textbf{d,} The instantaneous frequency was derived as the derivative of the phase versus time. The blue and red curves show the frequency without and with phase smoothing, respectively. The inset is a zoom in of the dashed box region.  
\textbf{e,} Retrieved frequency in the full sweep span. The frequencies were calibrated by 74 microcomb lines and were wrapped by $f_{r}$. The shaded box corresponds to the filtered pump and adjacent lines. 
%The heterodyne frequency results of a complete calibration signal containing 67 heterodyne signals. 
\textbf{f,} The retrieved frequencies were further unwrapped by the 50.08 GHz comb line spacing (dots). The unwrapped frequencies were fitted by a 9th-order polyfit. This fit was used to resample the time domain signal for ranging.}
\label{Extended_Fig2}
\end{centering}
\end{figure*}

\begin{figure*}[t]
\begin{centering}
\includegraphics[width=\linewidth]{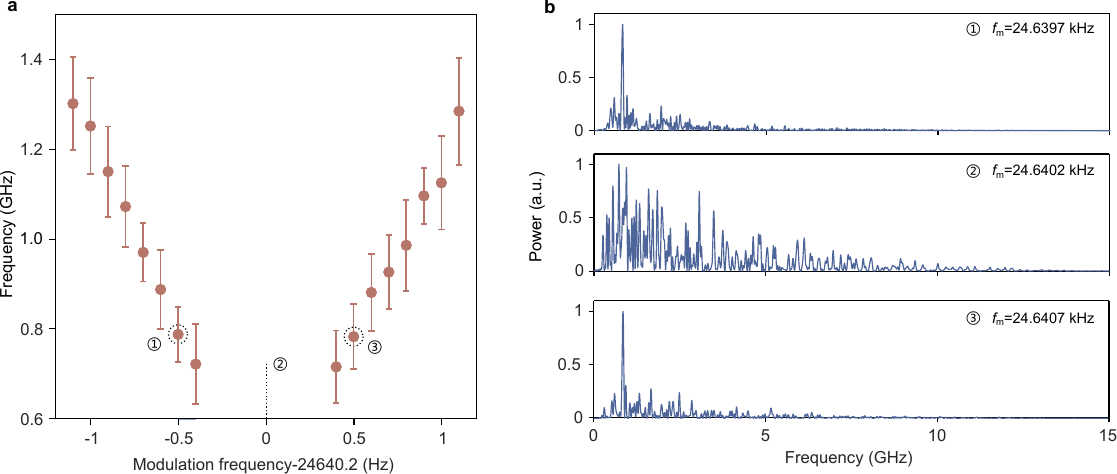}
\captionsetup{singlelinecheck=off, justification = RaggedRight}
\caption{{\bf Occurring frequency of the frequency spurs in the FDML laser.} \textbf{a,} Measured occurring frequency of the frequency spurs (abrupt frequency changes shown in Fig. \ref{Fig2}b) under different modulation frequencies. The frequency is about 1 GHz and is quite sensitive to the modulation frequency. The middle of the figure corresponds to a regime where there is no obvious occurring frequency. The error bar corresponds to the standard deviation of the occurring frequency in multiple measurements. \textbf{b,} Examples of the power spectra of the measured instantaneous frequency after removing the smoothed frequency change (e.g., frequency in the light curve minus the dark curve in Fig. \ref{Fig2}b). At a modulation frequency marked by 1 or 3 in panel \textbf{a}, the frequency spurs occur at a well-defined frequency. At a modulation frequency marked by 2, the power spectrum is broad, and there is no well-defined frequency.}
\label{Extended_Fig3}
\end{centering}
\end{figure*}

\begin{figure*}[t]
\begin{centering}
\includegraphics[width=\linewidth]{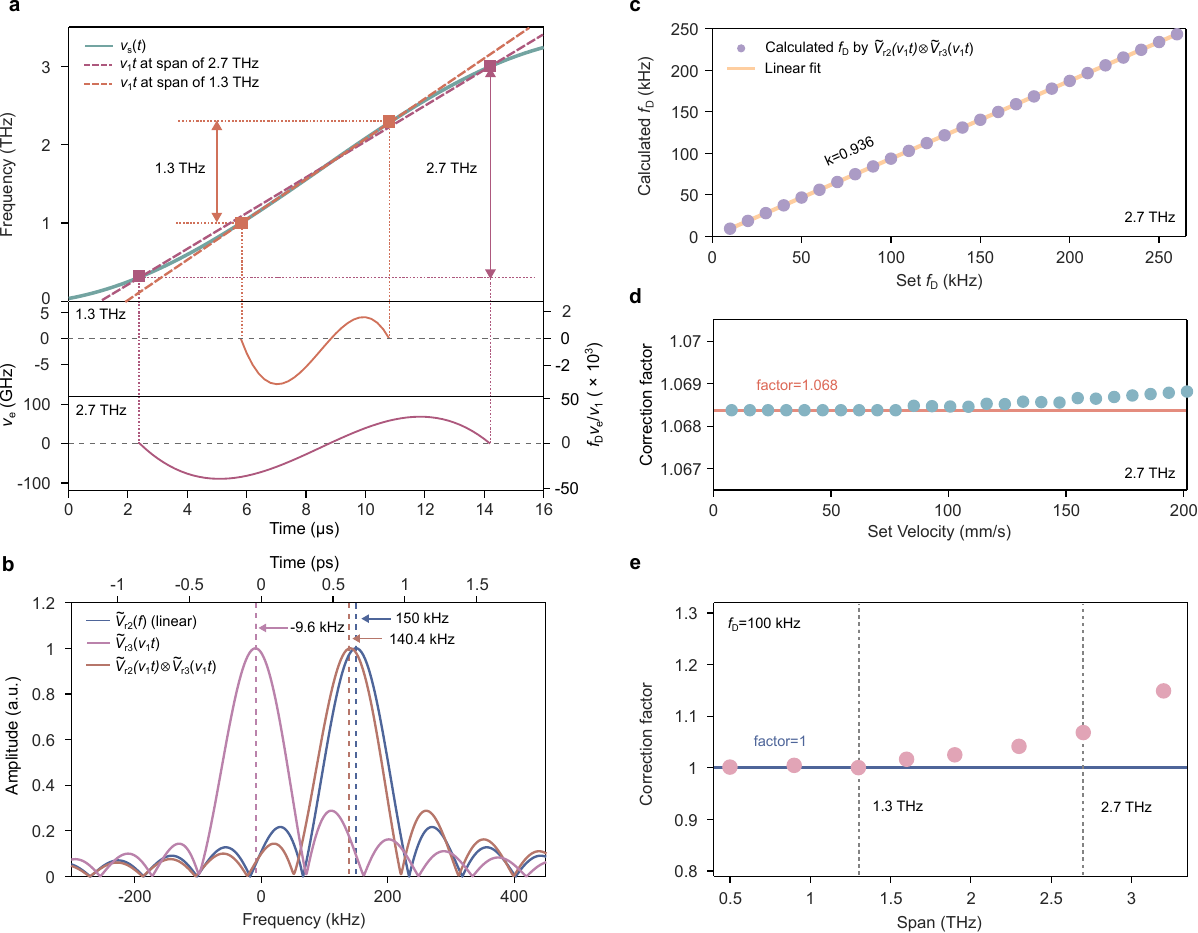}
\captionsetup{singlelinecheck=off, justification = RaggedRight}
\caption{{\bf Correction factor for velocity measurement with nonlinear frequency sweep.} \textbf{a,} The green curve is the 9th-order polyfit of the calibrated frequency sweep $\nu_{s}(t)$. We used linear fits $\nu_1 t$ to approximate the nonlinear frequency sweep, where $\nu_1$ is the overall chirp rate and was calculated as the frequency difference between the ending moment ($t$=$T$) and the starting moment ($t$=0) divided by the sweep time $T$ (see Methods). Since the frequency sweeps in a nonlinear way, there is an error frequency $\nu_{e}$ between $\nu_{s}$ and the linear fits $\nu_1 t$, which is plotted in the bottom panels. %\textcolor{magenta}{Even with the relatively small $\nu_e$ using the 1.3 THz bandwidth, transform-limited ranging spectra cannot be attained without frequency calibration and data resampling.} 
The error frequency $\nu_e$ becomes much larger when the 2.7 THz bandwidth was used.  
\textbf{b,} In a linear sweep case, the ranging spectrum has a peak centered at the set $f_{\rm D}$=150 kHz by Fourier transform the time domain data directly ($\widetilde{V}_{r2}(f)$ and dark blue curve). In a nonlinear sweep case, data resampling is needed, and $\widetilde{V}_{r3}(t)$ in Eq. \ref{eqnVTime} should be corrected to locate $f_{\rm D}$. By using the calibrated $\nu_{e}(t)$ for the 2.7 THz bandwidth shown in panel \textbf{a}, we can calculate $\widetilde{V}_{r3}(t)$ (thus, $\widetilde{V}_{r3}(\nu_1 t)$). The calculated  $\widetilde{V}_{r3}(\nu_1 t)$ corresponds to a peak at $-$9.6 kHz (light pink curve). Due to this shift,  the center of ranging spectrum calculated by $\widetilde{V}_{r2}(\nu_1 t)$$\otimes$$\widetilde{V}_{r3}(\nu_1 t) $ is located at 140.4 kHz (brick red curve), as opposed to the set 150 kHz. Thus, a correction factor $\alpha$=1.068 should be multiplied to the measured frequency shift to locate $f_{\rm D}$.  
\textbf{c,} Calculated $f_{\rm D}$ under different set velocity and $f_{\rm D}$ when using the $\nu_{e}(t)$ with 2.7 THz bandwidth in panel \textbf{a}. 
\textbf{d,} Correction factor $\alpha$ between the calculated and set $f_{\rm D}$ shown in panel \textbf{c}. 
\textbf{e,} Correction factor when selecting different optical bandwidths in $\nu_{s}(t)$ in panel \textbf{a} for FMCW measurements, assmuming a prior knowledge of $f_{\rm D}$=100 kHz.}
\label{Extended_Fig4}
\end{centering}
\end{figure*}

\setcounter{table}{0}
\renewcommand{\tablename}{\bf Extended Data Table}
\renewcommand\thetable{\arabic{table}}
%\FloatBarrier
\begin{table*}[h]
%\begin{centering}
\caption{\bf Detailed comparison with other FMCW LiDARs}
\label{Table_S1}
\centering
\begin{threeparttable}
\begin{tabular}{ p{0.6cm} p{1.8cm} p{1.5cm} p{2.6cm} p{2.6cm} p{1.85cm} p{2.0cm}p{2.7cm} p{0.9cm} }
\hline
 No. & Type & Resolution & \centering Normalized precision & \hspace{1em}Chirp rate & \centering Update rate & \centering Velocity uncertainty & \centering Vel. uncert./\par Chirp rate$^{\bf a}$ & Ref.\\
 \hline
1 & ECDL &  75 $\rm \mu m$  & \hspace{0.5em} 2.5 $ \rm \mu m \cdot \sqrt{s}$ & 1.0 $\times 10^{14}$ Hz/s &  \hspace{0.5em} 100 Hz & \hspace{0.5em} N/A & \hspace{0.3em} N/A &\cite{huang2024non}\\
2 & ECDL &  60 $\rm \mu m$ & \hspace{0.5em} 4.9 $ \rm \mu m \cdot \sqrt{s}$ & 1.25 $\times 10^{13}$ Hz/s & \hspace{0.5em} 5 Hz &  \hspace{0.5em} N/A & \hspace{0.3em} N/A& \cite{jia2021nonlinear}\\  
3 & ECDL &  60 $\rm \mu m$  & \hspace{0.5em} 0.21 $ \rm \mu m \cdot \sqrt{s}$ & 1.0 $\times 10^{13}$ Hz/s & \hspace{0.5em} 200 Hz & \hspace{0.5em}  N/A & \hspace{0.3em} N/A &\cite{zheng2022high}\\
4 & ECDL & 12 mm  & \hspace{0.5em} N/A & 6.25 $\times 10^{12}$ Hz/s & \hspace{0.5em} 500 Hz & \hspace{0.5em}  N/A & \hspace{0.3em} N/A & \cite{huang2022frequency}\\
5 & ECDL &  214 mm  & \hspace{0.5em} 1.3 $ \rm mm \cdot \sqrt{s}$ & 1.4 $\times 10^{12}$ Hz/s  & \hspace{0.5em} 1 kHz &  \hspace{0.5em} N/A & \hspace{0.3em} N/A& \cite{liu2024highly}\\
6 & DFB &  280 mm  & \hspace{0.5em}  N/A & 1.6 $\times 10^{13}$ Hz/s & \hspace{0.5em} 1.9 kHz & \hspace{0.5em} N/A  & \hspace{0.3em} N/A & \cite{martin2018photonic}\\% Antenna length of 50 mm
7 & DFB &  17 mm  & \hspace{0.5em} N/A & 1.1 $\times 10^{14}$ Hz/s  & \hspace{0.5em} 6.25 kHz & \hspace{0.5em} N/A & \hspace{0.3em} N/A & \cite{Wu_Nature2022large}\\
8 & DFB array &  \textbf{27} $\boldsymbol{\rm \mu m}^{\textbf{b}}$  & \hspace{0.5em} 0.25 $ \rm \mu m \cdot \sqrt{s}$ & 1.8 $\times 10^{15}$ Hz/s & \hspace{0.5em} 333 Hz & \hspace{0.5em}  N/A & \hspace{0.3em} N/A & \cite{dilazaro2018large}\\
9 & DBR &  1.2 mm & \hspace{0.5em} 17.9 $ \rm nm \cdot \sqrt{s}$ & 2.2 $\times 10^{16}$ Hz/s & \hspace{0.5em} 180 kHz & \hspace{0.5em} N/A & \hspace{0.3em} N/A & \cite{behroozpour2016electronic}\\
10 & DBR &  2.8 mm & \hspace{0.5em} 0.3 $ \rm \mu m \cdot \sqrt{s}$ & 7.5 $\times 10^{14}$ Hz/s & \hspace{0.5em} 16 kHz & \hspace{0.5em}  N/A & \hspace{0.3em} N/A & \cite{qian_NC2022video}\\
11 & PZT mod. & 125 mm  & \hspace{0.5em} N/A & 1.7 $\times 10^{15}$ Hz/s & \hspace{0.5em} 800 kHz & \hspace{0.5em} N/A & \hspace{0.3em} N/A & \cite{lihachev_NC2022low}\\
12 & PZT mod. & 35 mm  & \hspace{0.5em} 0.27 $ \rm mm \cdot \sqrt{s}$ & 8.4 $\times 10^{14}$ Hz/s & \hspace{0.5em} 10 kHz & \hspace{0.5em} N/A & \hspace{0.3em} N/A & \cite{tang2022hybrid}\\
13 & PZT mod. & 5.2 mm  & \hspace{0.5em} 1.1 $ \rm \mu m \cdot \sqrt{s}$ & 1.1 $\times 10^{13}$ Hz/s &  \hspace{0.5em} 80 Hz & \hspace{0.5em} \textbf{60} $ \boldsymbol{\rm \mu m/s}^{\bf c}$ & \hspace{0.3em} 5.4 $\mu$m/THz & \cite{wang2023laser}\\
14 & EO mod. & 59 mm  & \hspace{0.5em} 31.6 $ \rm \mu m \cdot \sqrt{s}$ & 4.4 $\times 10^{16}$ Hz/s & \hspace{0.5em} 10 MHz &  \hspace{0.5em} N/A & \hspace{0.3em} N/A & \cite{Kippenberg_Nature2020massively}\\
15 & EO mod. &  87 mm  & \hspace{0.5em} 4.9 $ \rm \mu m \cdot \sqrt{s}$ & 3.44 $\times 10^{16}$ Hz/s & \hspace{0.5em} 5 MHz &  \hspace{0.5em} 54 mm/s & \hspace{0.3em} 1.6 $\mu$m/THz & \cite{wang2024high}\\
16 & EO mod. & 125 mm  & \hspace{0.5em} N/A &  1.2 $\times 10^{16}$ Hz/s & \hspace{0.5em} 10 MHz &  \hspace{0.5em} N/A & \hspace{0.3em} N/A & \cite{snigirev_Nature2023ultrafast} \\
17 & EO mod. &  15.4 mm  & \hspace{0.5em} 5.6 $ \rm \mu m \cdot \sqrt{s}$ & 1.17 $\times 10^{16}$ Hz/s & \hspace{0.5em} 390 kHz &  \hspace{0.5em} N/A & \hspace{0.3em} N/A & \cite{he2023massively}\\
18 & EO mod. &  220 mm  & \hspace{0.5em} N/A & 1.0 $\times 10^{16}$ Hz/s & \hspace{0.5em} \textbf{20 MHz}$^{\bf d}$ & \hspace{0.5em} N/A & \hspace{0.3em} N/A & \cite{mrokon2024continuous}\\
19 & EO mod. &  80 mm  & \hspace{0.5em} N/A & 3.6 $\times 10^{14}$ Hz/s  & \hspace{0.5em} 100 kHz & \hspace{0.5em}  0.1 m/s & \hspace{0.3em} 277 $\mu$m/THz & \cite{lukashchuk2022dual}\\
20 & EO mod. &  25 mm  & \hspace{0.5em} N/A & 6.0 $\times 10^{13}$ Hz/s & \hspace{0.5em} 5 kHz & \hspace{0.5em}  7.8 mm/s & \hspace{0.3em} 130 $\mu$m/THz & \cite{dong2021frequency}\\
21 & EO mod. &  37.5 mm  & \hspace{0.5em} 74 $ \rm \mu m \cdot \sqrt{s}$ & 4.7 $\times 10^{12}$ Hz/s  & \hspace{0.5em} 588 Hz & \hspace{0.5em}  1 mm/s & \hspace{0.3em} 213 $\mu$m/THz & \cite{rogers2021universal}\\
22 & MEMS &  420 $\rm \mu m$ & \hspace{0.5em} N/A & 1.1 $\times 10^{17}$ Hz/s & \hspace{0.5em} 10 kHz &  \hspace{0.5em} N/A & \hspace{0.3em} N/A & \cite{okano2020swept}\\
23 & MEMS & 330 $\rm \mu m$ & \hspace{0.5em} 61 $ \rm nm \cdot \sqrt{s}$ & 1.7 $\times 10^{16}$ Hz/s & \hspace{0.5em} 10 kHz & \hspace{0.5em} N/A & \hspace{0.3em} N/A & \cite{hariyama2018high}\\
24 & MEMS &  3.5 mm & \hspace{0.5em} 22.4 $ \rm \mu m \cdot \sqrt{s}$ & 8.1 $\times 10^{15}$ Hz/s & \hspace{0.5em} 8 kHz & \hspace{0.5em} N/A & \hspace{0.3em} N/A & \cite{han2022high}\\
25 & MEMS &  130 $\rm \mu m$ & \hspace{0.5em} 2.5 $ \rm nm \cdot \sqrt{s}$ & 3.15 $\times 10^{15}$ Hz/s & \hspace{0.5em} 1 kHz & \hspace{0.5em} N/A & \hspace{0.3em} N/A &\cite{Newbury_OL2013comb}\\
26 & MEMS & 970 $\rm \mu m$ & \hspace{0.5em} N/A & 1.55 $\times 10^{15}$ Hz/s  & \hspace{0.5em} 5 kHz & \hspace{0.5em} 4 mm/s & \hspace{0.3em} 2.6 $\mu$m/THz & \cite{zhang2019laser}\\
27 & MEMS &  460 $\rm \mu m$ & \hspace{0.5em} 85.5 $ \rm nm \cdot \sqrt{s}$ & 1.4 $\times 10^{15}$ Hz/s  & \hspace{0.5em} 1 kHz & \hspace{0.5em} N/A & \hspace{0.3em} N/A & \cite{ula2019three}\\
28 & FDML &  55 ${\rm \mu m}$  & \hspace{0.5em} \textbf{0.27} $\boldsymbol{\rm nm \cdot \sqrt{s}}$ & \textbf{3.2} $\boldsymbol{\times 10^{17}}$ \textbf{Hz/s} & \hspace{0.5em} 24.6 kHz & \hspace{0.5em} 0.4 mm/s & \hspace{0.3em} \textbf{1.8 nm/THz} & This work\\
\hline
\end{tabular}

\begin{tablenotes}[l]
\item ECDL, external cavity diode laser; DFB, distributed feedback laser; DBR, distributed Bragg reflector laser; PZT mod., frequency tuning achieved via piezoelectric tuning; EO mod., frequency tuning achieved via electro-optic modulation; MEMS, the laser is micro-electro-mechanically tuned.

\item $^{\bf a}$Since velocity uncertainty is impacted by the overall chirp rate $\nu_1$ (Methods), this metric is also evaluated.

\item $^{\bf b}$Fine resolution achieved by stitching an array of 12 DFB lasers; a single DFB laser achieves a resolution of 333 $\mu$m.

\item $^{\bf c}$Low velocity measurement uncertainty achieved with low chirp rate.

\item $^{\bf d}$Frequency sweep range is 0.5 GHz.
\end{tablenotes}

\end{threeparttable}
\end{table*}

\end{document}